\documentclass[aip,pop,twocolumn]{revtex4}
\usepackage{CJK}
\usepackage{graphicx}
\usepackage{amsmath}
\usepackage{bm}

\begin{document}
\begin{CJK*}{GB}{}

\title{Agglomeration of microparticles in complex plasmas}
\author{Cheng-Ran Du}
\email{chengran.du@mpe.mpg.de}
\author{Hubertus M. Thomas}
\author{Alexei V. Ivlev}
\author{Uwe Konopka}
\author{Gregor E. Morfill}
\affiliation{Max Plank Institute for Extraterrestrial Physics, Garching 85741, Germany}

\begin{abstract}
Agglomeration of highly charged microparticles was observed and studied in complex plasma experiments carried out in a
capacitively coupled rf discharge. The agglomeration was caused by strong dust density waves triggered in a particle cloud
by decreasing neutral gas pressure. Using a high-speed camera during this unstable regime, it was possible to resolve the
motion of individual microparticles and to show that the relative velocities of some particles were sufficiently high to
overcome the mutual Coulomb repulsion and hence to result in agglomeration. After stabilising the cloud again through the
increase of the pressure, we were able to observe the aggregates directly with a long-distance microscope. We show
that the agglomeration rate deduced from our experiments is in good agreement with theoretical estimates. In addition, we
briefly discuss the mechanisms that can provide binding of highly charged microparticles in a plasma.
\end{abstract}

\maketitle
\end{CJK*}

\section{Introduction}

Dust agglomeration, also known as dust aggregation or dust coagulation, is a process of sticking dust particles together
after their collision \cite{Chokshi93,Blum06}. This process occurs in very different environments and plays an important
role in colloid chemistry \cite{Lin89}, astrophysics \cite{Wood98}, atmospheric physics \cite{Xiong01}, fusion physics
\cite{Winter98} as well as plasma physics \cite{Bouchoule}. Nano/microparticles trapped in a low-temperature plasma are
normally charged negatively due to high electron mobility \cite{Fortov05}, which results in the mutual repulsion between
them. Agglomeration of nanoparticles in plasmas -- the major mechanism operating during the so-called ``rapid growth'' (RG)
phase \cite{watanabe06} -- has been observed in numerous experiments \cite{Boufendi94,Nunomura08}. At the RG stage, most of
nanoparticles are neutral, some carry several electron charges, even fewer are charged positively \cite{Nunomura08,Ravi09},
and therefore the charge discreteness plays very important role. The agglomeration criteria for such particles were derived
theoretically, based on either the polarization-induced ion flux asymmetry model \cite{Mankelevich09} or the shadow
attraction due to neutral/plasma particle bombardment \cite{Bingham01}. The agglomeration of nanoparticles in a plasma has
been reported for sizes up to several hundreds of nm \cite{Hong06,Winter10}. As the size of particles immersed 
in a plasma increases, their charges become more and more negative (the charge is linearly
proportional to the particle size, for a one micron particle it ranges from several hundreds to several thousands of $e$,
depending on discharge conditions \cite{Fortov01, Fortov05,Schwabe07}). Therefore, agglomeration of microparticles in a
plasma has been believed so far to be practically impossible due to strong Coulomb repulsion. This process was primarily
studied (both experimentally and theoretically) in a neutral environment \cite{Ivlev02,Konopka05}, showing that positive and
negative charges carried by individual particles (in overall neutral ensemble) can enormously accelerate the agglomeration.

In this article we report on a series of experiments where dust-acoustic (density) waves trigger agglomeration of
microparticles. We show that the agglomeration can occurs under typical discharge conditions, for particles of different
sizes and materials, and the aggregates formed in a plasma can be observed {\it in situ}, using a long-distance microscope.

\section{Experiments and observations}

The experiments were performed on the ground in the PK-3 Plus chamber \cite{Thomas08}, which was originally designed for microgravity
experiments onboard the International Space Station (ISS). The chamber (which consists of two parallel electrodes driven in push-pull mode
at a frequency of $13.56$ MHz) was placed horizontally and single microparticles (no agglomerates) were injected into the discharge from a
special dispenser. A vertical laser sheet (perpendicular to the electrodes) with a thickness of about $100~\mu$m was used to illuminate the
particles in the plasma, the scattered light was recorded at $90^{\circ}$ by a high-speed CCD camera. A detailed description of the PK-3
Plus setup can be found in \cite{Thomas08}.

\begin{figure*}
\includegraphics[width=0.7\textwidth]{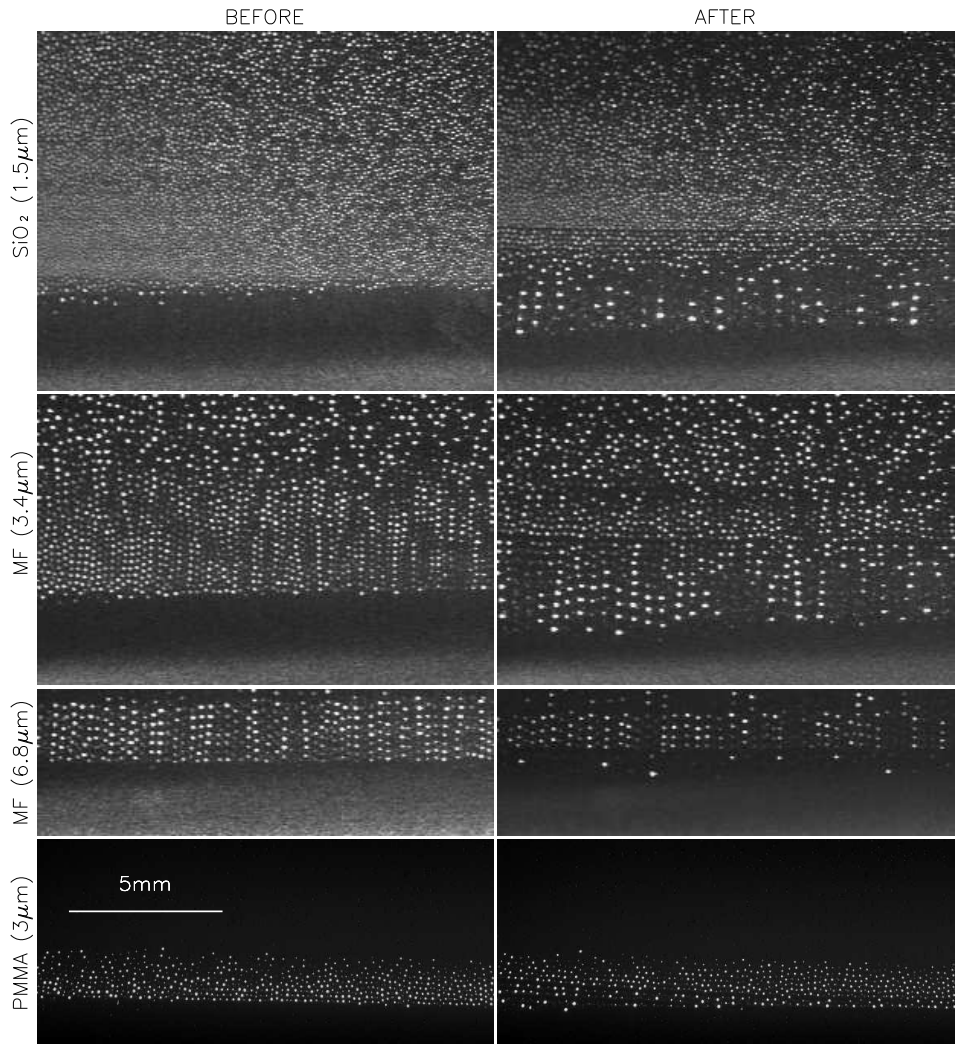}
\caption{Particle cloud before the dust density waves were triggered (first column) and after the waves were stopped (second
column). Shown are results of the experiments with different particles: The first row represents SiO$_2$ particles of $1.5~\mu$m
diameter, the second row -- MF particles of $3.4~\mu$m diameter, the third row -- MF particles of $6.8~\mu$m diameter, and the
fourth row -- PMMA particles of $3~\mu$m diameter. \label{figure1}}
\end{figure*}

Microparticles were first injected into argon plasma (typically maintained at a rf peak-to-peak voltage of $\simeq25$~V and gas pressure of
$\simeq50$~Pa), to form a stable cloud above the bottom electrode. Then, in order to accelerate particles, dust-acoustic (density) waves
were triggered in the cloud by decreasing gas pressure well below the instability threshold \cite{Schwabe07} (down to 10--20~Pa). This
unstable regime was maintained for several minutes, and then the gas pressure was restored to the initial (stable) level. Such experiments
were performed with particles of different sizes and materials.

We compared profiles of the dust clouds before and after the wave excitation, as illustrated in Fig.~\ref{figure1}. After
the excitation, we observed aggregates levitated at the bottom of the cloud (they can be identified as brighter spots having
larger distances to their neighbors). For SiO$_2$ particles of $1.5~\mu$m diameter (first row), multiple layers of
aggregates (apparently of different sizes) were seen after the unstable regime. The aggregates were separated from single
particles by a narrow but well-seen gap, while before the wave excitation there were only a few bigger particles near the
lower edge. The similar situation was also observed for melamine-formaldehyde (MF) particles of $3.4~\mu$m and $6.8~\mu$m
diameter (second and third row, respectively) as well as for polymethyl-methacrylate (PMMA) particles of $3~\mu$m diameter
(fourth row).

\begin{figure}
\includegraphics[width=0.35\textwidth]{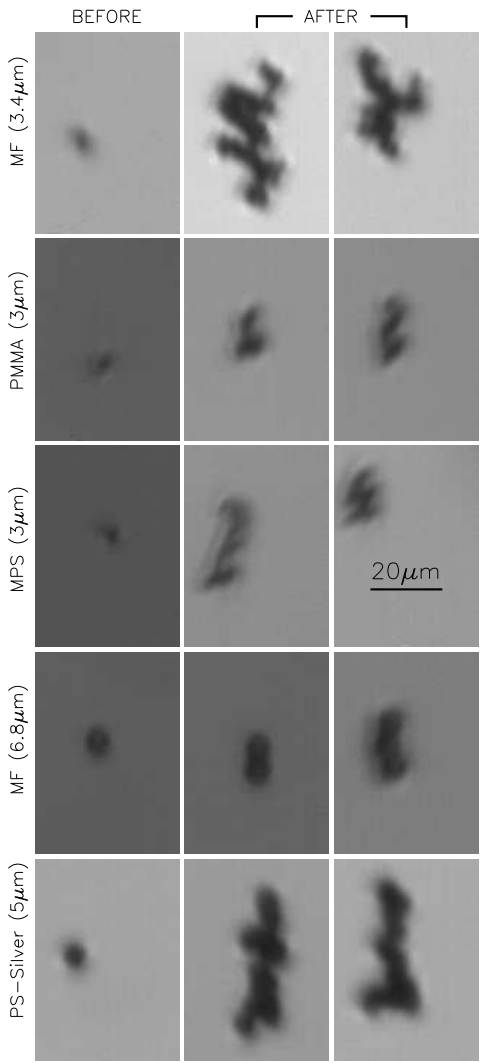}
\caption{Single microparticles before the waves were triggered (first column) and examples of aggregates after the waves were
stopped (second and third column), all observed with a long-distance microscope. Experiments were performed with different
particles, the first row shows MF particles of $3.4~\mu$m diameter, the second row -- PMMA particles of $3~\mu$m
diameter, the third row -- MPS particles of $3~\mu$m diameter, the fourth row -- MF particles of $6.8~\mu$m diameter,
and the fifth row -- silver-coated PS particles of $5~\mu$m diameter. \label{figure2}}
\end{figure}

In order to take a close look at the structure of the aggregates we used a long-distance microscope (with a diffusive light
source from the opposite side of the chamber), which allowed us to perform {\it in situ} observations. Due to limited
resolution of the microscope, we were only able to analyze the experiments with sufficiently large particles. The results
are presented in Fig.~\ref{figure2}: In the experiments with MF, PMMA, and ``magnetic polystyrene'' \cite{footnote} (MPS)
particles of about $3~\mu$m diameter (first, second, and third row, respectively) the aggregates were usually composed of
several particles and had rather irregular (often elongated) structure. For larger MF particles of $6.8~\mu$m diameter
(fourth row) the aggregates were typically formed of two-three particles. It is noteworthy that for silver-coated
polystyrene (PS) particles of $5~\mu$m diameter (fifth row) the agglomeration was also possible and in fact very effective,
resulting in the formation of relatively large elongated aggregates. This strongly suggests that the agglomeration of
microparticles in a plasma is a quite generic phenomenon occurring both for dielectric and conductive particles.

\section{Analysis}

\begin{figure}[b!]
\includegraphics[width=0.45\textwidth]{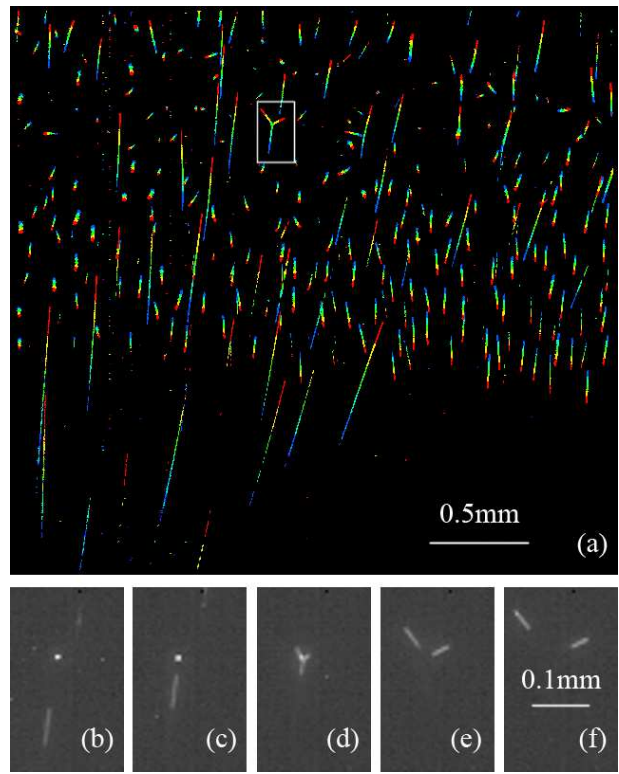}
\caption{Trajectories of individual microparticles in the wave. (a) The trajectories are depicted by superimposing five
consecutive video frames, color-coded from dark blue to red. (b)-(f) Example of a two-particle scattering collision [magnified
rectangular box from (a)], the scattering is shown as a sequence of five consecutive frames. The
results are for MF particles of $2.5~\mu$m diameter.
\label{figure3}}
\end{figure}

The agglomeration in a plasma can only be achieved if heavily charged particles acquire the kinetic energy which is
sufficiently high to overcome the interparticle repulsion. In our experiment, the particles were accelerated by the
dust-acoustic waves (which were presumably triggered by the ion streaming instability \cite{Fortov05}). In order to track
the individual particles inside the wave, we performed a series of experiments with MF particles of $2.5~\mu$m diameter and
recorded their motion during two seconds. We used a high-speed camera at 2000 frames per second (fps) with full exposure,
recording the field of view of 3~mm$\times$3~mm (at a spatial resolution of 210 pixel per mm). The waves propagated mostly
vertically, with an average wave period of about $60$~ms (at a gas pressure of $20$~Pa). The waves were highly nonlinear,
with a minor fraction of particles moving in one direction at rather high velocities, while most of them were moving slowly
in the opposite direction. To visualize the particle trajectories, five consecutive video frames were color-coded and
overlapped, as shown in Fig.~\ref{figure3}a.

\begin{figure}
\includegraphics[width=0.3\textwidth]{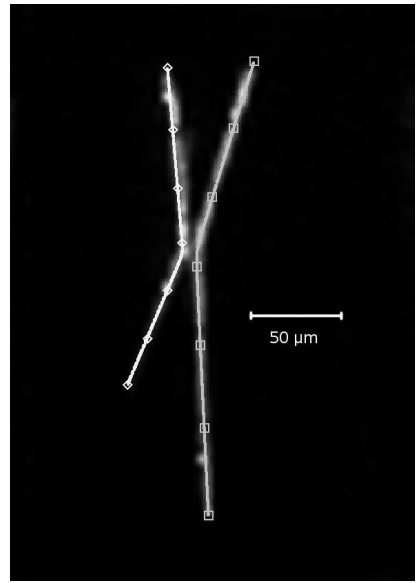}
\caption{Comparison of a two-particle scattering collision with the simulations. The simulated trajectories (marked by
diamonds and squares) are overlapped with the observed particle trajectories (from the experiment shown
in Fig.~\ref{figure3}). One of the trajectories (diamonds) becomes noticeably dimmer at one end, suggesting that the scattering
plane (in the center-of-mass reference frame) is not parallel to the illuminating laser sheet. \label{figure4}}
\end{figure}

In order to analyze the agglomeration process in detail, we first have to evaluate the parameters that determine the
interparticle interaction. Obviously, the most critical parameter for such analysis is the charge of microparticles -- if it
is too high the coagulation is inhibited. A simple estimate of the charge from the orbit-motion-limited (OML) theory is just
not possible for our conditions, since the plasma parameters in dense particle clouds are poorly known. As an alternative,
we were able to derive the charge from rarely occurring {\it scattering} collisions of two particles. We supposed that the
observed collision events are representative, in the sense that charges of the colliding particles are close to the mean
particle charge in the whole cloud. An example of the collision event is shown in Fig.~\ref{figure3}b-f. For the analysis we
assumed the Debye-H\"{u}ckel interaction potential which depends on two unknown parameters: the charge $Q$ and the screening
length $\lambda$. These parameters were derived by simulating the pair collisions and fitting the resulting trajectories to
the observed ones. We took into account neutral gas friction \cite{Epstein24} and used the particle velocities and
coordinates measured before the collision as the initial conditions. Since the velocities and coordinates were naturally
measured with some errors, the fittings of $Q$ and $\lambda$ was performed by varying the initial conditions within the
estimated error bars. The best fit was then defined by maximizing the correlation between the simulated and measured
trajectories, which is illustrated in Fig.~\ref{figure4}. We found that the fit is practically insensible to the value of
$\lambda$, apparently because the relative velocities of colliding particles were so high that the scattering mostly
occurred at distances smaller than the screening length. This observation significantly simplified the subsequent analysis
and allowed us to directly employ the Rutherford's scattering theory \cite{Schwabe07,Konopka}. Given the fact that the
typical collision event is much shorter than the gas damping timescale, the charge can be obtained from the formula
$Q=-\sqrt{2Eb\tan\frac12\theta}$, where $E=\frac14mv^2$ is the (initial) kinetic energy of the relative particle motion
(where $m$ is the mass of a single particle and $v$ is the relative velocity), $\theta$ is the scattering angle, and $b$ is
the impact parameter for colliding particles. This yields the charge $Q=-(1650\pm170)~e$, where the uncertainty is given by
the standard deviation averaged over ten collision events. However, one should keep in mind that the actual uncertainty is
somewhat larger, because the collision plane (in the center-of-mass reference frame) is never perfectly parallel to the
illuminating laser sheet (which has a finite thickness), as one can see from Fig.~\ref{figure4}.

\begin{figure}
\includegraphics[width=0.4\textwidth]{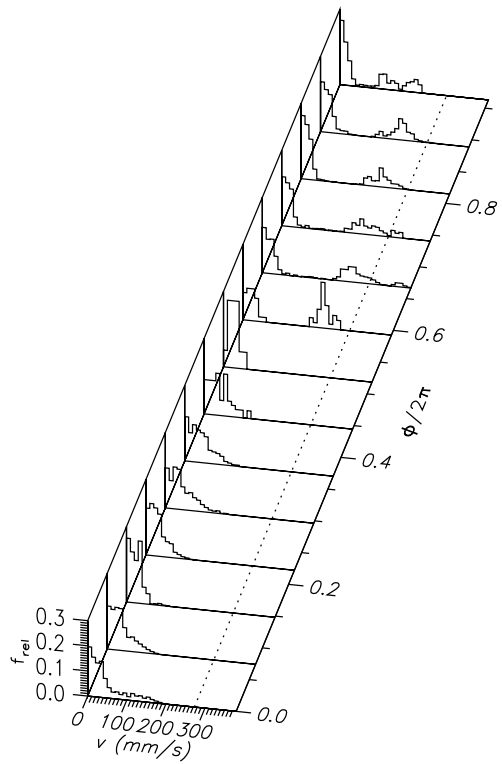}
\caption{Relative velocity distribution of microparticles, $f_{\rm rel}(v)$, for different phases $\phi$ of the wave period.
The distribution is normalized to unity, the dash line corresponds to the critical velocity $v_{\rm cr}(=273$~mm/s). \label{figure5}}
\end{figure}

Two colliding particles can touch each other only if their relative velocity exceeds a critical (threshold) value determined
by the Coulomb repulsion. Using the same assumptions as above, we readily obtain the critical velocity,
\begin{equation}\label{v_cr}
v_{\rm cr}=\frac{|Q|}{\sqrt{\frac12ma}},
\end{equation}
which has the following scaling on the particle size: $v_{\rm cr}\propto a^{-1}$ (here we also assume $Q\propto a$). For the
experiment shown in Fig.~\ref{figure3} we get $v_{\rm cr}\simeq273$ mm/s. In order to evaluate the probability for particles
in the cloud to reach the critical velocity, we tracked the individual particles manually and estimated their velocities
from the lengths of the trajectories at each video frame. From this analysis we deduced the {\it relative velocity}
distribution (normalized to unity), $f_{\rm rel}(v)$, which is shown in Fig.~\ref{figure5} for different phases $\phi$ of
the wave. This plot clearly demonstrates that for certain $\phi$ there is a small fraction of particles with the relative
velocities exceeding $v_{\rm cr}$.

Using these results, we estimated the agglomeration rate for pair collisions (per unit volume, assuming that the sticking
probability upon the collision is unity),
\begin{equation}
  R=\frac{1}{T}\int_0^{T}\mathrm{d}t\;{n^2}\int_{0}^{\infty}\mathrm{d}v\;{\sigma}{v}f_{\rm rel},
  \label{agglomeration rate}
\end{equation}
where $n(t)$ is the (time-dependent) particle number density in the cloud, $T$ is the wave period, and the collision cross
section,
\begin{displaymath}
 \sigma(Q,v)=
 \left\{
  \begin{array}{ll}
   0,                                    & v\leq{v_{\rm cr}}; \\
   4\pi a^2\left(1-v_{\rm cr}^2/v^2\right), & v>{v_{\rm cr}},
  \end{array}
 \right.
\end{displaymath}
is determined by the critical velocity given by Eq.~(\ref{v_cr}). The calculations yield $R\simeq0.8$~mm$^{-3}$s$^{-1}$ for
the experiment illustrated in Figs~\ref{figure3}-\ref{figure5}. In order to compare these estimates with the experiment, we
selected a part of the visible cloud (assuming the horizontal homogeneity) which represent a slice of the thickness
$\simeq0.1$~mm (about the thickness of the illuminating laser sheet), width of $3$ mm, and height of $5.8$ mm (full height
of the cloud). The number of (double) aggregates predicted by Eq. (\ref{agglomeration rate}) to appear in this volume after
five minutes of the unstable regime is $\simeq450$, which is 5--6 times larger than the number of aggregates (82) we
actually observed. Taking into account that the sticking probability of the colliding particles can be substantially smaller
than unity and also that the aggregates in the considered experiment consisted (on average) of more than two particles, we
conclude that the theory provides reasonable agreement with the experiment.

\section{Discussion and conclusion}

Let us discuss the mechanisms that can effectively bind particles in a plasma. If we assume a homogeneous charge
distribution over the spherical particle surfaces, the repulsive Coulomb force $F_{\rm C}$ between them is equivalent to
that between two point charges $Q$ located at the centers of the spheres. Therefore, the force at the contact,
\begin{equation}\label{F_C}
  F_{\rm C}=\frac{Q^2}{4a^2},
\end{equation}
does not depend on $a$. On the other hand, the charge at one particle induces a dipole moment at another particle. This
results in the attractive force which depends on the permittivity of the particle material and also scales as
$\propto(Q/a)^2$, thus tending to balance $F_{\rm C}$.

However, there are (at least) two other very efficient attraction mechanisms that can overcome the repulsive force
(\ref{F_C}). One of them is due to the van der Waals interactions between individual molecules of microparticles. For two
identical spherical particles, Hamaker \cite{Hamaker37} gives the following expression of the attractive force:
\begin{equation}\label{F_vdW}
  F_{\rm vdW}=\frac{A_{\rm H}a}{12\delta^2},
\end{equation}
where $A_{\rm H}$ is the material-dependent Hamaker constant and $\delta(\ll a)$ is the effective distance between surfaces
of two particle at the contact \cite{Hamaker37,Blum06}. (If the polarization of spheres is neglected then $A_{\rm
H}=\pi^2\rho^2\alpha$, where $\rho$ is the number density of molecules and $\alpha$ is the corresponding London-van der
Waals constant.) By comparing Eqs (\ref{F_C}) and (\ref{F_vdW}) we see that the van der Waals attraction is stronger than
the Coulomb repulsion when $A_{\rm H}>Q^2\delta^2/a^3(\propto a^{-1})$. Let us consider an aggregate composed of two SiO$_2$
particles ($A_{\rm H}=6.5\times10^{-20}$~J) of the diameter $1.5~\mu$m (the smallest particles used in the experiments, see
the first row of Fig.~\ref{figure1}). Assuming that the particles carry the charge $Q=-1000~e$ each and are separation by
$\delta\simeq3$~{\AA} at the contact, we obtain that $F_{\rm vdW}$ is two orders of magnitude stronger than $F_{\rm C}$.
Note that for metal particles (such as Au, Ag, Cu) $A_{\rm H}$ is typically in the range $(25-40)\times10^{-20}$~J
\cite{Israelachvili}, and therefore the binding force for them should be even stronger. One should also point out that the
particle coating (see, e.g., the fifth row of Fig.~\ref{figure2}) may affect van der Waals attraction, provided the coating
thickness is comparable to or exceeds $\delta$ and the $\alpha$ constant of the coating material is different from that of
the particle material.

The second mechanism of attraction can be associated with the discreteness of the charge distribution over the particle
surface (here, we do not discuss the physics of the electron states at dielectric surfaces). The ``mean
distance'' between the electrons at the surface, which can be estimated as $\Delta_e\sim a\sqrt{4\pi e/|Q|}\propto a^{1/2}$,
is typically of a few dozens of nm. Therefore, when particles get in contact their interaction should strongly depend on
whether there are individual electrons in a close proximity of the contact spot. For instance, if there is one electron {\it
at each} particle surface near the contact spot, this will exert an additional repulsive force which can be as large as
$\sim e^2/\delta^2$. However, if an electron is only attached to {\it one} of the particles, this results in the additional
{\it attraction} between them due to the local polarization of the particle material. Simple calculations (representing each
particle as an infinite half-space filled with a material of high complex permittivity) yield the following attractive force
due to the ``charge discreteness'':
\begin{equation}\label{F_disc}
F_{\rm disc}=\frac{e^2}{4\delta^2}.
\end{equation}
By comparing Eq. (\ref{F_disc}) with Eq. (\ref{F_C}) we obtain that the charge discreteness force should overcome the
Coulomb repulsion when $(a/\delta)^2>(Q/e)^2$, and this condition does not depend on $a$. For SiO$_2$ particles considered
above (to illustrate the van der Waals attraction), $F_{\rm disc}$ is an order of magnitude stronger than $F_{\rm C}$ and
hence can also contribute to the binding. Note that $F_{\rm disc}$ is really of short-range and can only be important when
particles get very close to each other ($\delta\lesssim\Delta_e$) -- otherwise the interaction is described by usual
``mean-field'' formulas. Also, $F_{\rm disc}$ must be strongly dependent on the particle (surface) conductivity and
presumably be much larger for dielectric particles than for conductive (or coated) ones.

One should make a common remark concerning both binding mechanisms discussed above. In our calculations we used rather crude
assumption that the charges of individual particles in a plasma and charges of particles bound in a cluster are the same.
This assumption is quite justified as the ``first approach'', since the estimated binding forces turn out to be much
stronger than the Coulomb repulsion. However, for a more careful analysis of the problem one should take into account the
mutual influence of the neighboring particles on their charges (see, e.g., \cite{Manweiler93,Vladimirov03}). In fact, the
binding force should strongly depend not only on the magnitude of the particle charge, but also on the charge distribution
over the particle surface. This very complex issue undoubtedly requires further careful analysis.

To summarize, by triggering dust-acoustic waves in a cloud of microparticles suspended in the pre-sheath region of a low
pressure rf discharge (which is referred to as the unstable regime), we were able to accelerate the particles to velocities
which were sufficiently high to overcome their mutual Coulomb repulsion. This resulted in the agglomeration -- the formation
of clusters consisting of two, three, or more particles -- yet this effect was observed for particles of different sizes and
materials. By tracking individual particles and resolving their scattering collisions with a high-speed camera, we calculated
the mean particle charge and reconstructed the relative velocity distribution function. This allowed us to estimate the
coagulation rate during the unstable regime and compare it with the experimental observations. Additionally, by using a
long-distance microscope we were able to observe the aggregates {\it in situ}, levitated in a plasma after the wave
excitation.

The results presented in this paper unambiguously demonstrate that microparticles, even exposed in a plasma environment, can
form stable clusters due to strong short-range binding forces. We believe that these results can be of great importance to
various fields ranging from plasma processing (where this effect can be utilized, e.g., to remove dust contamination) to
astrophysics (where the effect can have a strong impact on the coagulation of cosmic dust).

\begin{acknowledgments}
The authors would like to acknowledge the valuable discussions with M. Schwabe, S. Shimizu, and T. Shimizu as well as the technical
support from T. Hagl. This research was funded by " Die Raumfahrt-Agentur des Deutschen Zentrums f\"{u}r Luft und Raumfahrt e. V. 
mit Mitteln des Bundesministeriums f\"{u}r Wirtschaft und Technologie aufgrund eines Beschlusses des Deutschen Bundestages 
unter dem F\"{o}rderkennzeichen 50 WP 0203".
\end{acknowledgments}


\begin{thebibliography}{28}
\bibitem[{foo()}]{footnote}
\bibinfo{note}{We used microspheres with an even dispersion of magnetic
  material (gamma Fe$_2$O$_3$ and Fe$_3$O$_4$) throughout the PS bead. The
  particles contain 20~\% iron and were super-paramagnetic (i.e., with no
  residual magnetism).}
\end{thebibliography}
\end{document}